\documentclass[doublecol]{epl2}

\def\bra#1{\mathinner{\langle{#1}|}}
\def\ket#1{\mathinner{|{#1}\rangle}}

\title{The importance of inter-site coherences in the x-ray absorption spectra of mixed-valent systems}
\shorttitle{The importance of inter-site coherences in the x-ray absorption} 

\author{Subhra Sen Gupta\footnote{E-mail : subhra@phas.ubc.ca} \and Hiroki Wadati\footnote{Present address : Department of Applied Physics and
QPEC, University of Tokyo, Hongo,
Tokyo 113-8656, Japan.} \and G. A. Sawatzky}
\shortauthor{Subhra Sen Gupta \etal}

\institute{
 Department of Physics and
Astronomy, University of British Columbia, Vancouver, BC V6T 1Z1,
Canada.}

\pacs{78.70.Dm}{X-ray absorption spectra}
\pacs{75.30.Mb}{Valence fluctuation, Kondo lattice, and heavy-fermion phenomena }
\pacs{78.20.Bh}{Theory, models, and numerical simulation}

\abstract{We study the importance of inter-site coherences and
quantum interference effects in the x-ray absorption spectroscopy
(XAS) of mixed-valent systems. We demonstrate its importance, first
for a simple diatomic mixed-valent molecule to elucidate the physics
involved, and finally for mixed-valent Ti oxides, using a model
calculation including full spin and orbital multiplicities. Our
calculations demonstrate the inefficacy of conventional
approaches of describing the XAS from mixed-valent systems
as incoherent combinations of XAS from the pure-valent end members,
and that multiplets forbidden in the single-site approximation can
be reached within a multi-site description. These conclusions play
an important role in the interpretation of XAS and
resonant x-ray scattering (RXS) in terms of the electronic structure
of strongly correlated systems.}

\begin{document}

\maketitle

\section{Introduction}Strongly correlated electron systems, of which the transition metal
(TM) compounds are exemplary, form one of the central themes of
present day condensed matter physics. An important feature of these
systems is the relatively small spatial extent of their valence $3d$
wave functions, which form narrow bands of widths often matched or
overwhelmed by the electron-electron Coulomb interactions, leading
to strong correlation effects. The physical properties of these
systems are largely determined by rather local physics dictated by
point group symmetries, Hund's rule coupling of the $d$ electrons
and short-range superexchange interactions. XAS along with x-ray
magnetic circular (XMCD) and linear (XMLD) dichroism, involving
$2p\rightarrow 3d$ core-valence transitions on the TM site, has
served as an important tool for investigating this local physics,
\emph{viz.} the oxidation states, the local orbital occupancies,
spin-state, crystal field symmetry, metal-ligand hybridization,
Coulomb correlation effects \emph{etc.}, in these
systems~\cite{degroot-kotani-book,original-papers-XAS-XMCD}.

Furthermore, recently developed RXS methods where the cross section
can be expressed in terms of simple first-order fundamental XAS
spectra~\cite{Maurits-RIXS-REXS-condmat}, can, at least in
principle, provide unique information concerning the ordering of
spin, charge and orbitals in these systems. However, the information
contained in the x-ray energy dependent intensities at superlattice
reflections often cannot be accessed, especially for doped systems,
because of disagreement with conventionally used single-site cluster
like model calculations of the fundamental XAS spectra. In fact this
disagreement may point to the importance of intersite coherences
playing an important role.

\begin{figure}
\begin{center}
\includegraphics[angle=0,width=0.40\textwidth]{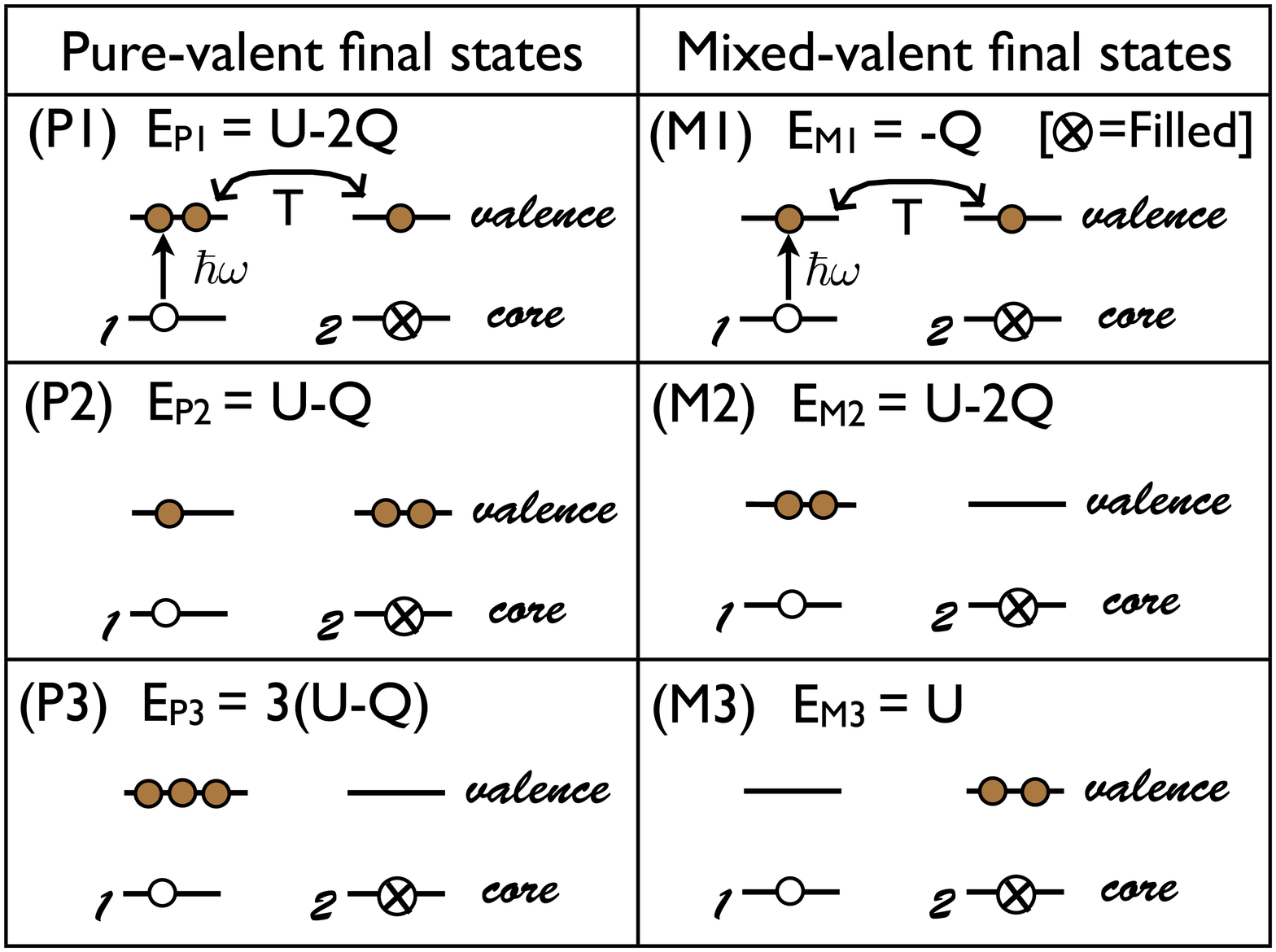}
\caption {(\emph{colour online}) Schematic representing the various
relevant final states and their energies, for the XAS of {\em pure-valent}
[$(P1)-(P3)$] and {\em mixed-valent} [$(M1)-(M3)$] correlated systems, in a
two-site picture. $U=$ valence electron-electron Coulomb repulsion,
$Q=$ core-hole, valence-electron attraction.}\label{simple-scheme}
\end{center}
\vspace{-0.7cm}
\end{figure}

Given the complexity of the problem due to several competing
interactions, the inference of ground state properties from the
spectral lineshapes requires detailed many-body crystal or ligand
field multiplet calculations within a configuration interaction (CI)
approach~\cite{degroot-kotani-book}. The success of such local
approaches, involving a \emph{single central correlated TM site}
coordinated by a shell of ligand atoms (\emph{e.g.,} oxygen, sulphur
\emph{etc.}), implicitly assumes a \emph{pure-valent system} with an
integral \emph{nominal} filling per TM site (\emph{i.e.}, before
hybridization with the ligands), which ensures that : (1) in the
ground state, charge fluctuations involving neighboring TM sites are
suppressed by the Hubbard $U$; and \emph{more importantly}, (2)
the locally photo-excited electron in the valence $d$ orbital is
strongly bound to the core-hole on the same atom in the final state,
by an attractive $2p$-$3d$ Coulomb potential, $Q$ ($\sim 1.2U$).
This strongly suppresses charge fluctuations involving neighbouring
TM sites, since the inter-TM-site effective hopping integrals, $T$,
are usually much smaller than
$Q$~\cite{Bocquet-late-TM,Bocquet-early-TM}. See
fig.~\ref{simple-scheme} (configurations $P1-P3$) for a pictorial
representation of the relevant physics. The net outcome is that the
core-hole and extra valence electron form core-excitonic bound
states, so that a single-site description suffices.

However, things become more interesting for \emph{mixed-valent} or
\emph{valence fluctuating} systems~\cite{mixed-valence-Varma}. Many
of the exotic manifestations of strong correlation \emph{e.g.},
various kinds of charge and spin density wave states, colossal
magnetoresistance, high temperature superconductivity \emph{etc.},
arise when the parent compound is hole or electron doped, so that
the average nominal number of electrons per TM site is no longer an
integer. The interpretation of XAS in these systems is a very
important problem, as it can yield important information concerning
quantum fluctuations involving also neighboring TM sites, which in
turn would serve to explain the observed exotic properties. The
general assumption, till now, has been that one can get away with
\emph{incoherently averaging the spectra (weights determined by the
doping) calculated on the basis of single TM site clusters with
integral nominal fillings}~\cite{HY-Huang-Nature}. But this approach
neglects crucial quantum interference effects arising out of
coherent charge fluctuations between different TM sites and may miss
some important physics. It implies a phase-separated scenario
instead of uniform mixed-valency. It is the conditions under which
these coherent mixing effects play an important role, that we
address in this paper.

To treat a doped system, the minimum extension should involve two TM
sites with either an \emph{odd} (mixed-valent) or an \emph{even}
(pure-valent) total number of $d$ electrons, in the dimer, in the
starting state. One can then compare the \emph{``exact"} calculation
for such a two-site cluster with the incoherent sum of the component
spectra from single-site clusters, to verify the importance of
coherence effects. Fig.~\ref{simple-scheme} compares the basic
energetics of the various states involved in the presence of a
core-hole on one of the atoms and an additional valence electron,
between mixed-valent and pure-valent systems. This demonstrates that
while the energy difference between the two lowest energy final
states, which are connected by a hopping integral $T$, is $Q$ for
the pure-valent case, it is only $|U-Q|$ for the mixed-valent one.
It is easy to see then that for $T$ much smaller than $U$ and $Q$,
the mixing of final states shown for the pure-valent case is small
and can mostly be neglected. However, for $T\sim |U-Q|$, the
mixing in the mixed-valent case can be very large indeed.

\section{A simple model} 
\begin{figure}
\begin{center}
\vspace{-0.2cm}
\includegraphics[angle=0,width=0.49\textwidth]{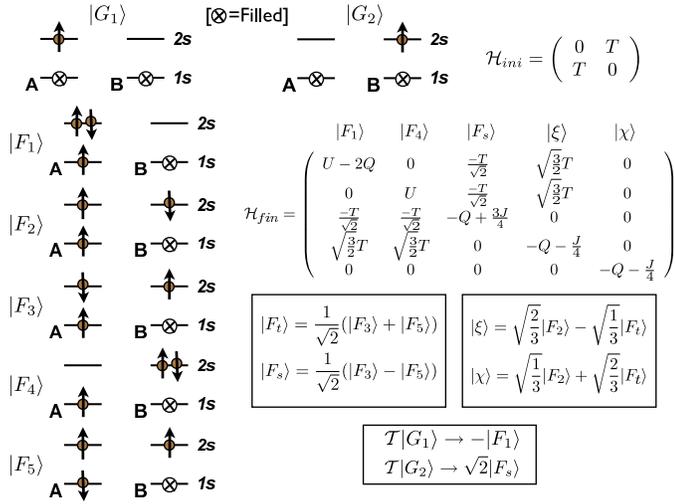}
\caption {(\emph{colour online}) Schematic representation of the
various relevant \emph{initial} ($\ket{G_{1}}-\ket{G_{2}}$) and
\emph{final} ($\ket{F_{1}}-\ket{F_{5}}$) configurations (basis
states) for the Li$_2^+$ molecule, shown along with the initial
(${\cal H}_{ini}$) and final (${\cal H}_{fin}$) state Hamiltonians.
The final state Hamiltonian is written in terms of a transformed
basis which is defined below it. The action of the transition
operator (${\cal T}$) on the initial basis states is also
listed.}\label{hamiltable}
\end{center}
\vspace{-0.7cm}
\end{figure}

To describe the physics involved we start with a simple model
system, not complicated by a lot of multiplet structure \emph{viz.},
a fictitious diatomic Li$_2^+$ molecule with orbitally
non-degenerate (but including spin degeneracies), core $1s$ and
valence $2s$ levels at each Li site, as shown in
fig.~\ref{hamiltable}. This basic model was used previously to
describe core-level photoemission from mixed-valent systems,
illustrating there the importance of the broken symmetry
ansatz.~\cite{Sawatzky-Lenselink}. The core $1s$ levels are filled
with a total of 4 electrons, and the valence $2s$ levels share one
electron in all, in the initial state, corresponding to 50\% doping
of the Li$_2$ molecule. We calculate the $1s\rightarrow 2s$
excitation spectrum for this system, which contains the same basic
physics as an allowed core-to-valence transition in XAS.

The Hamiltonian for this molecule is given by :
$${\cal H}=\sum_{i,\sigma}\epsilon_{1s}c_{1s}^{i\sigma\dagger}c_{1s}^{i\sigma} +
\sum_{i,\sigma}\epsilon_{2s}c_{2s}^{i\sigma\dagger}c_{2s}^{i\sigma}+
\sum_{\sigma}(Tc_{2s}^{A\sigma\dagger}c_{2s}^{B\sigma} + h.c.)$$
\vspace{-0.38cm}
\begin{equation}
-J\sum_{i}\overrightarrow{\sigma_{i}}\cdot\overrightarrow{S_{i}}
-Q\sum_{i,\sigma}(n_{h}^{1s})^{i}c_{2s}^{i\sigma\dagger}c_{2s}^{i\sigma}
+U\sum_{i}n_{2s}^{i\uparrow}n_{2s}^{i\downarrow} \label{Hamil}
\end{equation}
where, the sum over $i=A,B$ runs over the two Li sites, and
$\sigma=\uparrow,\downarrow$ sums over the two spin projections.
$c_{1s,2s}^{i\sigma\dagger}$ ($c_{1s,2s}^{i\sigma}$) creates
(destroys) an electron at the site $i$ with spin $\sigma$, in the
$(1s,2s)$ orbitals, respectively. Above,
$n_{2s}^{i\sigma}=c_{2s}^{i\sigma\dagger}c_{2s}^{i\sigma}$ is the
number operator for the $2s$ electrons,
$(n_{h}^{1s})^{i}=(2-\sum_{\sigma}c_{1s}^{i\sigma\dagger}c_{1s}^{i\sigma})$
counts the number of holes in the core $1s$ level, while
($\overrightarrow{S_{i}}$) and ($\overrightarrow{\sigma_{i}}$)
denote the operators for the core and the valence spins, all at site
$i$. In eq.~(\ref{Hamil}) the first two terms refer to the on-site
energies of the two orbitals ($\epsilon_{2s}>\epsilon_{1s}$), which
merely shift the transition energy and which we take to be zero.
The subsequent terms describe, respectively, the $2s$ electron
inter-site hopping $T$, the Hund's rule like $1s-2s$ ferromangetic
exchange $J$, the $1s$-hole, $2s$-electron on site attraction $Q$,
and the $2s-2s$ on-site coulomb repulsion $U$. In our convention
while $T<0$, $J$, $Q$ and $U$ are \emph{positive numbers}.

In fig.~\ref{hamiltable} we show the basis configurations involved
in the initial and final states, their diagonal energies and the
off-diagonal matrix elements resulting from eq.~(\ref{Hamil}), for
the initial (${\cal H}_{ini}$) and final (${\cal H}_{fin}$)
Hamiltonians. The initial Hilbert space involves two degenerate
configurations $\ket{G_{1}}$ and $\ket{G_{2}}$, and the $2s$
hopping, $T$, is the only off-diagonal term. We take the $2s$
electron to be \emph{spin-up}. This stabilizes the bonding state,
\emph{i.e.}, $\ket{\psi_{g}} =
\frac{1}{\sqrt{2}}(\ket{G_{1}}+\ket{G_{2}})$, as the ground state
for $T<0$.
The final Hilbert space is more complex, consisting of five relevant
configurations $\ket{F_{i}}$ ($i=1,5$) as shown in
fig.~\ref{hamiltable}. While $\ket{F_{1}}$ corresponds to
\emph{neutral} Li (Li$^{0}$), the configurations $\ket{F_{2}}$ to
$\ket{F_{5}}$ are obtained from the Li$^{+}$ ion, at the core-hole
site $A$. The off-diagonal matrix elements now also involve $J$. We
have excluded a state composed of the two \emph{spin-down} $2s$
electrons and a \emph{spin-up} $1s$ electron, as it does not connect
to any of the other $\ket{F_{i}}$ configurations. It is instructive
to talk in terms of spin states at the core-hole site $A$ ($S_{A}$),
as well as the total spin $S$ including the effect of the
\emph{spectator spin doublet} at site-$B$. In this sense,
$\ket{F_{1}}$ and $\ket{F_{4}}$ are $\ket{S=1/2, m=+1/2}$ states
derived from the parents $\ket{S_{A}=1/2, m_{A}=+1/2}$ and
$\ket{S_{B}=0, m_{B}=0}$. These occur at the energies $(U-2Q)$ and
$U$. The state $\ket{F_{2}}$, at energy $(-Q-\frac{J}{4})$, consists
of the triplet state $\ket{S_{A}=1, m_{A}=+1}$ at site-$A$ and the
doublet component $\ket{S_{B}=1/2, m_{B}=-1/2}$ at site-$B$.
As demonstrated in fig.~\ref{hamiltable}, we combine $\ket{F_{3}}$
and $\ket{F_{5}}$ to form singlet ($\ket{F_{s}}$) and triplet
($\ket{F_{t}}$) eigenstates of $S_{A}$. Both $\ket{F_{s}}$ with
$\ket{S_{A}=0, m_{A}=0}$ at $(-Q+\frac{3J}{4})$,
and $\ket{F_{t}}$ with $\ket{S_{A}=1, m_{A}=0}$ at
$(-Q-\frac{J}{4})$ can combine with the $B$-site spin to form
\emph{total spin doublets}. $\ket{F_{2}}$ being actually the
$m_{A}=+1$ component of the site-$A$ triplet, is degenerate with
$\ket{F_{t}}$. As shown fig.~\ref{hamiltable}, these two degenerate
triplets can be combined, at $T=0$, into eigenstates of the total
spin $S$, to yield the \emph{doublet} $\ket{\xi}=\ket{S=1/2,
m=+1/2}$ and the \emph{quartet} $\ket{\chi}=\ket{S=3/2, m=+1/2}$, both at $(-Q-\frac{J}{4})$.

The final state Hamiltonian, ${\cal H}_{fin}$, written in this new basis
$\{\ket{F_{1}},\ket{F_{4}},\ket{F_{s}},\ket{\xi},\ket{\chi}\}$, now
involves only $T$ in its off-diagonal elements, as shown
in fig.~\ref{hamiltable}. This is diagonalized to yield the final
eigenstates $\ket{\psi_{f}^{k}}$ ($k=1,5$). The transition operator
for the $1s\to 2s$ excitation is given by~:
\begin{equation}
{\cal T}={\cal
M}\sum_{\sigma}c_{2s}^{A\sigma\dagger}c_{1s}^{A\sigma}
\label{transition}
\end{equation}
with a spin-independent transition matrix element ${\cal M}$. The
action of ${\cal T}$ on each of the initial basis states is
enumerated in fig.~\ref{hamiltable}, where the negative signs arise
because of the antisymmetric nature of the basis configurations, and
results in transitions to only those final states that involve a
singlet or doublet component at site-$A$.
Without loss of generality we set ${\cal M}=1$. Then the excitation
spectrum is given by the expression~:
\begin{equation}
I(\omega)=\sum_{k}|\bra{\psi_{f}^{k}}{\cal
T}\ket{\psi_{g}}|^{2}\delta(\omega-(E_{f}^{k}-E_{g}))
\label{specfunc}
\end{equation}

\begin{figure}
\begin{center}
\includegraphics[angle=0,width=0.30\textwidth]{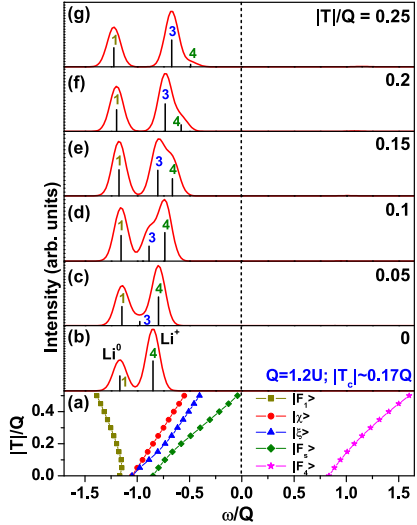}
\vspace{-0.2cm} \caption {(\emph{colour online}) (a) Evolution of the
scaled transition energies, $\omega/Q$, to the final eigenstates
with the magnitude of the scaled $2s$-$2s$ hybridization, $|T|/Q$,
for the Li$_2^+$ model. The states at $T=0$ from which each
eigenstate evolves are marked; (b)-(g) evolution of the
$1s\rightarrow 2s$ excitation spectrum, with increasing $|T|/Q$,
within the same model. Numbers against peaks refer to the final eigenstate from which they originate.}\label{Li-spec}
\end{center}
\vspace{-0.95cm}
\end{figure}

The evolution of the transition energies,
$\omega=(E_{f}^{k}-E_{g})=(E_{f}^{k}-T)$, with $T$ is shown in
fig.~\ref{Li-spec} (a), for the values $U=0.83Q$ (\emph{i.e.},
$Q=1.2U$), $J=0.2Q$. \emph{All numerical values for eigenenergies
($E$), hoppings ($T$) \emph{etc.} for this model are henceforth
quoted in units of $Q$.} At $T=0$ these correspond to the diagonal
energies of ${\cal H}_{fin}$ as discussed above. Now both
$\ket{F_{s}}$ and $\ket{F_{t}}$ hybridize with
$\{\ket{F_{1}},\ket{F_{4}}\}$ with an amplitude
$-\frac{T}{\sqrt{2}}$, while $\ket{F_{2}}$ hybridizes with the same
states with an amplitude $T$. This, together with the definitions in
fig.~\ref{hamiltable}, shows that while the doublet $\ket{\xi}$
hybridizes with $\{\ket{F_{1}},\ket{F_{4}}\}$ with an amplitude
$\sqrt{\frac{3}{2}}T$, the quartet state $\ket{\chi}$ is completely
non-bonding. This explains why the state $\ket{\chi}$ does not
disperse at all with $T$, which shows up as a linear movement of the
transition energy, $\omega$, in fig.~\ref{Li-spec} (a), due to the
hybridization shift, $T$, of the ground state, while $\ket{\xi}$
disperses only weakly being repelled in opposite directions by
$\ket{F_{1}}$ and $\ket{F_{4}}$.

Now turning to the trends in the intensities, we find that for
$T=0$, the spectra can be visualized as an incoherent combination of
spectra from Li$^{0}$, which involves a \emph{doublet} to
\emph{doublet} transition ($\ket{G_{1}}\rightarrow \ket{F_{1}}$) at
$(U-2Q)$, and from Li$^{+}$ which starts from a \emph{singlet} at
site-$A$ ($\ket{G_{2}}$) and hence a transition to only the
\emph{local singlet}, $\ket{F_{s}}$, at $(-Q+\frac{3J}{4})$ is
possible for $T=0$, the \emph{triplet} $\ket{F_{t}}$ being forbidden
(fig.~\ref{Li-spec} (b)). These peaks appear in the intensity ratio
of $1:2$, corresponding to just one way of exciting the Li$^{0}$,
compared to the two ways of exciting the Li$^{+}$ ion. For small
values of $T$, till about $T=0.05$ as is evident from fig.~\ref{Li-spec}
(b-c), the spectrum basically still has only two peaks with very
little intensity in other regions. However, as $T$ increases, there is spectral weight
transfer from higher to lower energy so that the two peaks tend
towards equal intensities. This could lead
to an incorrect conclusion about the actual $2s$ electron count 
and, in the case of doped systems, result in incorrect estimates of doping.
We also see that the energy splitting between the Li$^{0}$ and
Li$^{+}$ derived states shows very little change up to a critical
value of $|T_{c}|\sim|U-Q|\sim 0.17$ or as long as $|T|<|U-Q|$, and
changes almost linearly for $|T|>|U-Q|$. An analysis of these
spectra in terms of an incoherent sum of the parent ion spectra
would obviously lead to incorrect conclusions regarding both the
composition of the material as well as the ``chemical" shifts of the
core-hole spectra.

Inspection of the unbroadened intensities for finite $T$, especially
for $T=0.1$, $0.15$ and $0.2$ (fig.~\ref{Li-spec} (d)-(f)), reveals the
other crucial fact that for any realistic $T$, \emph{transitions are
now possible to previously forbidden (at $T=0$) multiplets in the
Li$^{+}$-like region, which now shows two peaks with finite
intensities that change systematically.} This can be understood on
the basis that at finite $T$ only the total spin $S$ is relevant.
Thus starting from the ground state doublet $\ket{S=1/2, m=+1/2}$,
we can only reach final state doublets, which are the states derived
from $\ket{F_{1}}$, $\ket{F_{4}}$, $\ket{F_{s}}$ and $\ket{\xi}$,
while the quartet $\ket{\chi}$ which does not mix, is still
forbidden. So $\ket{\xi}$, which has 0.33 of the
$A$-site forbidden triplet $\ket{F_{t}}$ mixed in at $T=0$, always
has a finite contribution from it for finite $T$, demonstrating how
forbidden multiplets in a single-site representation now appear as
extra structures in the more correct multi-site approach.

\section{Realistic calculations for Ti oxides} Although this simple model clarifies the basic physics, more
realistic calculations including the full orbital and spin
degeneracies of a TM ion, multiplet interactions and charge transfer
from ligand atoms \emph{etc.}, are needed to approach real
correlated systems. To this end we have calculated the isotropic XAS
from a \emph{mixed-valent} [Ti$-$O$-$Ti] cluster, mimicking the
half-doped oxide La$_{0.5}$Sr$_{0.5}$TiO$_{3}$. The pure-valent end
members SrTiO$_{3}$ (STO) and LaTiO$_{3}$ (LTO) consist of Ti$^{4+}$
($3d^{0}$) and Ti$^{3+}$ ($3d^{1}$) ions, respectively, so that the
mixed-valent compound has one electron per two Ti atoms. Accordingly
our cluster nominally has one electron shared between the two Ti
atoms. Alternately, such a mixed-valent bond could also arise at a
STO/LTO band-insulator/Mott insulator heterostructure interface, and
calculations similar to the present one could help to understand the
observed correlated metallic state~\cite{HY-Huang-Nature,STO-LTO} at
the interface. Although the use of only a dimer of Ti atoms is a
severe approximation to real systems in which there is no
dimerization, a calculation with the full coordination of O and
neighbouring Ti atoms is at the limits of feasability even with
modern computers. However, in many real systems in fact, dimer
formation seems to occur in particular phases such as in the low
temperature insulating phase of VO$_{2}$~\cite{VO2-dimer}, the Zener
polaron like scenarios proposed for some
Manganites~\cite{Zener-polaron}, Peierls or spin-Peierls like ground
states in quasi one-dimensional systems, to name a few.

Our model calculations include the TM core $2p$ and valence $3d$
states as well as the O $2p$ valence states. Full multiplet Coulomb
interactions and spin-orbit interaction within the Ti $3d$ manifold,
the spin-orbit interaction within the Ti core $2p$, and final state
core-valence ($2p$-$3d$) multiplet interactions on the Ti were taken
into account. The relevant spin-orbit parameters and the multipole
Coulomb Slater-Condon parameters, $(F_{dd}^{2},F_{dd}^{4})$ and
$(F_{pd}^{2},G_{pd}^{1},G_{pd}^{3})$ were obtained from Cowan's
atomic Hartree-Fock code~\cite{Cowan}, and following the usual
convention for TM ions~\cite{antonides} the Slater-Condon parameters
were reduced to 80\% of their atomic values. Based on reported
values in the literature~\cite{Bocquet-early-TM} the multiplet
averaged $3d$-$3d$ and $2p$-$3d$ Coulomb interactions were set to
$U=4.5$ eV and $Q=1.2U=5.4$ eV, so that $|U-Q|=0.9$ eV. The Ti
$3d-$O $2p$ charge transfer energy ($\Delta$) was defined with
respect to a $d^{1}$ system as in LTO, (\emph{i.e.},
$\Delta=E(d^{2}\underline{L})-E(d^{1})$, where $\underline{L}$
denotes a ligand-hole state) and set to $\Delta=6.0$
eV~\cite{Bocquet-early-TM}. An octahedral ($O_{h}$) crystal field of
magnitude $10Dq=1.5$ eV was used on both the Ti sites to effectively
take into account the octahedral coordination around the Ti ions,
not included in this calculation. The metal-ligand hopping is given
by the Slater-Koster parameters $(pd\sigma,pd\pi)$, where
$pd\pi=-pd\sigma/2.2$ was used in all cases~\cite{Bocquet-early-TM}.
Since our model includes only one ligand oxygen atom explicitly
shared between the two Ti atoms, it is quite likely to find more
than one hole on the oxygen atom. Thus the usual reasoning that the
density of holes on the oxygen atoms is very low and hence the
Hubbard $U$ on the O $2p$ is ineffective, no longer holds. So we
have also taken into account the full multiplet Coulomb interaction
in the O $2p$ manifold with the multiplet averaged interaction
$U_{pp}=4.0$ eV~\cite{Tjeng-Oxy-Auger}, while $F_{pp}^{2}$ was again
obtained from Cowan's code. Calculations were performed for various
degrees of covalency, starting from a realistic value of
$pd\sigma=-2.5$ eV and gradually reducing it through \{-1.5, -0.5,
-0.1\} eV to the free ion limit. For the purpose of truncation of
the basis set, the convergence of the spectrum was checked as a
function of the number of ligand-hole excitations for the largest
value of hopping ($pd\sigma=-2.5$ eV), retaining up to 4 ligand-hole
($\underline{L}^{4}$) excitations. The spectrum was found to
converge well at $\underline{L}^{3}$. Hence the basis was truncated
at $\underline{L}^{3}$ for all the calculations, the ones for lower
hopping being expected to converge even sooner.
\begin{figure}
\begin{center}
\includegraphics[angle=0,width=0.47\textwidth]{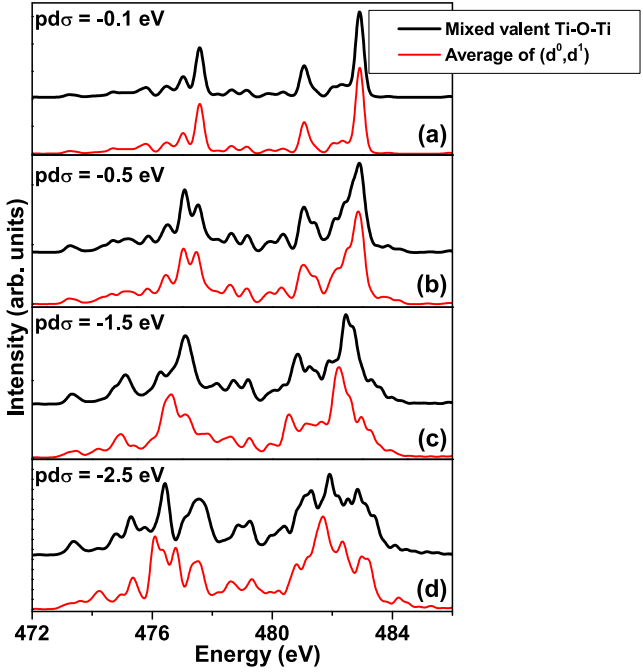}
\caption {(\emph{colour online}) Calculated XAS from a
\emph{mixed-valent} [Ti$-$O$-$Ti] cluster (\emph{thick line})
(mimicking La$_{0.5}$Sr$_{0.5}$TiO$_{3}$), compared with the
incoherently averaged XAS (\emph{thin line}) from pure-valent
$d^{0}$ (mimicking SrTiO$_{3}$ : Ti$^{4+}$) and $d^{1}$ (mimicking
LaTiO$_{3}$ : Ti$^{3+}$) [Ti$-$O] clusters, as a function of the
metal-ligand hopping : (a) $pd\sigma$=-0.1 eV, (b) $pd\sigma$=-0.5
eV, (c) $pd\sigma$=-1.5 eV, (d) $pd\sigma$=-2.5 eV (realistic).
As the hopping is increased from very small (a) to realistic (d) values,
the mixed-valent spectrum deviates farther from its incoherent
counterpart.}\label{Spec-mixed}
\end{center}
\vspace{-0.8cm}
\end{figure}

These spectra are to be compared to an incoherent average of
single-site pure-valent calculations, to bring out the differences.
However an atomic $O_{h}$ crystal field multiplet calculation from a
Ti$^{3+}$ and a Ti$^{4+}$ ion does not suffice here, because in
choosing a [Ti$-$O$-$Ti] cluster, we broke the $O_{h}$ symmetry
around the Ti ion via the Ti$-$O bond, and hence we can only compare
to a single-site calculation that also breaks the symmetry in the
same way. This is achieved by performing calculations for a
[Ti$^{3+}$$-$O] and a [Ti$^{4+}$$-$O] cluster, retaining the
$10Dq=1.5$ eV $O_{h}$ crystal field, again for the various hoppings
as mentioned above. All of the parameters, other than $\Delta$, were
kept the same as for the mixed-valent cluster. However, whereas
$\Delta=6.0$ eV was used for the [Ti$^{3+}$$-$O] cluster ($3d^{1}$),
an appropriately compensated value of $\Delta'=\Delta-U=1.5$ eV
($\Delta'=E(d^{1}\underline{L})-E(d^{0})$) was used for the
[Ti$^{4+}$$-$O] cluster ($3d^{0}$), which is the effective charge
transfer energy to the $d^{0}$ site within the coherent two-site
mixed-valent calculation. For these calculations the full basis set
was used without any truncation. A small Gaussian broadening of FWHM=0.3 eV was used in all cases.

In fig. \ref{Spec-mixed} (a)-(d) we show the XAS spectra for the
mixed-valent cluster (thick black line), starting with the smallest
hopping ($pd\sigma=-0.1$ eV, atomic-like situation) in panel (a) and
then gradually increasing it to the realistic value of
$pd\sigma=-2.5$ eV in panel (d). In each case we compare it with the
spectrum obtained by averaging (thin red line) the XAS from the
pure-valent [Ti$^{4+}$$-$O] and [Ti$^{3+}$$-$O] clusters. Just as in
our Li$_{2}^{+}$ calculation, we find that the case for the smallest
hopping (a) has almost no coherence between the Ti sites and is
formally equivalent to the incoherent average of the XAS from the
single site clusters representing the pure-valent end members.
However, as we increase the hopping, coherence begins to develop
between the two Ti sites and the XAS from the mixed-valent cluster
begins to depart from the corresponding incoherent average. While
for $pd\sigma=-1.5$ eV the differences are quite noticeable, for the
more realistic $pd\sigma=-2.5$ eV, they become truly pronounced. So
the single site approximation completely breaks down for realistic
hoppings in a mixed-valent system.
\begin{figure}
\begin{center}
\includegraphics[angle=0,width=0.4\textwidth]{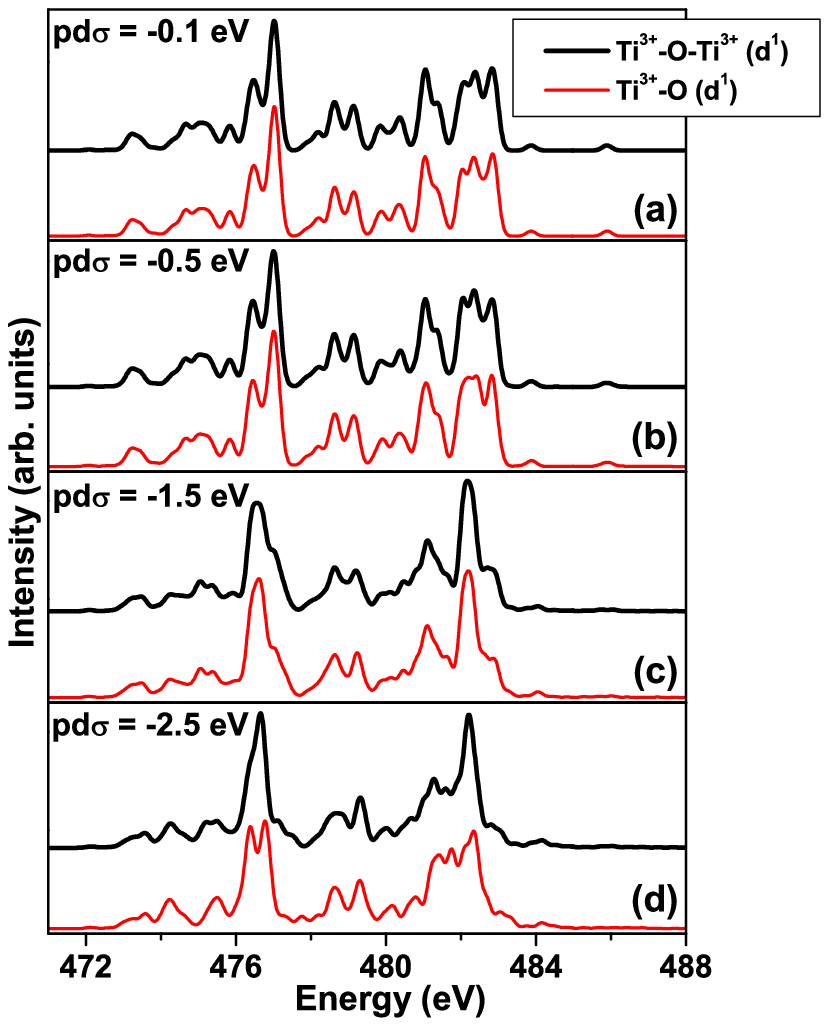}
\caption {(\emph{colour online}) Calculated XAS from a pure-valent ($d^{1}$)
[Ti$^{3+}$$-$O$-$Ti$^{3+}$] cluster (\emph{thick line}) (mimicking LaTiO$_{3}$) compared with that from a pure-valent
[Ti$^{3+}$$-$O] cluster (\emph{thin line}), as a function of the
metal-ligand hopping : (a) $pd\sigma$=-0.1 eV, (b) $pd\sigma$=-0.5
eV, (c) $pd\sigma$=-1.5 eV, (d) $pd\sigma$=-2.5 eV (realistic).
Except for the largest hopping
($pd\sigma$=-2.5 eV) (d), the spectra are nearly identical for all other cases. Even for
$pd\sigma$=-2.5 eV, the differences are much smaller than that for
the mixed-valent case (\emph{c.f.}, fig. \ref{Spec-mixed}).}
\label{Spec-pure-d1}
\end{center}
\vspace{-0.8cm}
\end{figure}

\begin{figure}
\begin{center}
\includegraphics[angle=0,width=0.4\textwidth]{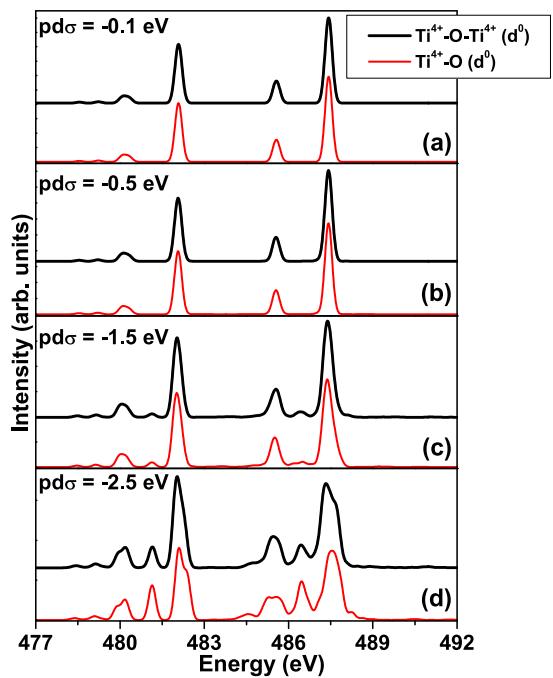}
\vspace{-0.0cm} \caption {(\emph{colour online}) Calculated XAS from
a pure-valent  ($d^{0}$) [Ti$^{4+}$$-$O$-$Ti$^{4+}$] cluster (\emph{thick
line}) (mimicking SrTiO$_{3}$~\cite{no-match-expt}) compared with that from a
pure-valent [Ti$^{4+}$$-$O] cluster (\emph{thin line}), as a
function of the metal-ligand hopping : (a) $pd\sigma$=-0.1 eV, (b)
$pd\sigma$=-0.5 eV, (c) $pd\sigma$=-1.5 eV, (d) $pd\sigma$=-2.5 eV
(realistic). Except for the largest hopping ($pd\sigma$=-2.5 eV) (d),  the spectra are nearly identical for all
other cases. Even for $pd\sigma$=-2.5 eV, the differences are much
smaller than that for the mixed-valent case (\emph{c.f.}, fig.
\ref{Spec-mixed}).} \label{Spec-pure-d0}
\end{center}
\vspace{-0.8cm}
\end{figure}

Finally, we would like to demonstrate that this problem is much more
pronounced in mixed-valent systems compared to pure-valent ones. For
the latter, the differences between the two approaches is relatively
small, so that one can get away with a single site approximation. To
this end we have calculated also the XAS from the two-site
pure-valent clusters [Ti$^{4+}$$-$O$-$Ti$^{4+}$] and
[Ti$^{3+}$$-$O$-$Ti$^{3+}$] which are meant to represent the
stoichiometric compounds STO and LTO. In accordance with reported
values for STO and LTO~\cite{Bocquet-early-TM}, $\Delta=6.0$ eV was
used for both calculations. While $\Delta=6.0$ eV is a bit on the
higher side for STO, the exact value is unimportant and this serves
well to bring out the physics involved~\cite{no-match-expt}. All other parameters were
exactly the same as used for the mixed-valent cluster. These were
compared to the XAS from the pure-valent single-site clusters
calculated for the same parameter values. The results are shown, for
the Ti$^{3+}$ ($d^{1}$) and Ti$^{4+}$ ($d^{0}$) calculations, in
fig. \ref{Spec-pure-d1} (a)-(d), and fig. \ref{Spec-pure-d0}
(a)-(d), respectively, for the same values of hopping, and shown in
the same order as in fig. \ref{Spec-mixed}. In each panel the XAS
from the \emph{two-site pure-valent} (thick black line) and
\emph{single-site pure-valent} (thin red line) clusters are
compared. We find that the only really noticeable differences exist
for the largest hopping of $pd\sigma=-2.5$ eV (panels (d) in each
case), which are much smaller than in the mixed-valent case (fig.
\ref{Spec-mixed} (d)). 

\section{Conclusion} Our calculations demonstrate the importance of inter TM site coherences in the XAS
of mixed-valent (doped) strongly correlated systems. These effects
cause non-trivial differences between the XAS from a truly
mixed-valent system and that obtained by incoherently averaging XAS
from the corresponding pure-valent end members, whenever the hybridization is comparable to the
\emph{difference} between the valence band Coulomb repulsion ($U$)
and the core-hole potential ($Q$). This results in strong
changes in energy shifts, and a dramatic redistribution of spectral weight away from the statistical limit, between the so-called pure-valent
components. It also shows that multiplets forbidden in the single-site
approximation can effectively be reached within a multi-site
description, for realistic hoppings. For pure-valent systems, this
problem is much less severe and the single impurity approximation
works reasonably well. This physics has important consequences for
doped correlated systems, in general, and TM oxide heterostructures
(like LTO/STO) having interfaces across which a change of nominal
valency of the TM ion occurs~\cite{HY-Huang-Nature,STO-LTO}.
This can also play an important role in distinguishing
between charge disproportionation and Zener polaron or dimeron like
scenarios~\cite{Zener-polaron}.

\acknowledgments
We would like to acknowledge funding from the
Canadian agencies NSERC, CFI and CIFAR. SSG would like to thank
Prof. D. D. Sarma for the permission to modify and use the
spectroscopy codes developed in his group.

\end{document}